\documentclass{ws-ijmpcs}

\begin{document}

\markboth{MANEL PERUCHO}
{JET STABILITY, DYNAMICS AND ENERGY TRANSPORT}

%
\catchline{}{}{}{}{}
%

\title{JET STABILITY, DYNAMICS AND ENERGY TRANSPORT}

\author{MANEL PERUCHO}

\address{Departament d'Astronomia i Astrof\'{\i}sica, Universitat de Val\`encia, C/ Dr. Moliner 50\\
Burjassot, Valencian Country, 46100, Spain\\
manel.perucho@uv.es}

\maketitle

\begin{history}
\received{Day Month Year}
\revised{Day Month Year}
\end{history}

\begin{abstract}
Relativistic jets carry energy and particles from compact to very large scales compared with their initial radius. 
This is possible due to their remarkable collimation despite their intrinsic unstable nature. In this contribution, 
I review the state-of-the-art of our knowledge on instabilities growing in those jets and several stabilising 
mechanisms that may give an answer to the question of the stability of jets. In particular, during the last years we have 
learned that the limit imposed by the speed of light sets a maximum amplitude to the instabilities, contrary to the 
case of classical jets. On top of this stabilising mechanism, the fast growth of unstable modes with small wavelengths 
prevents the total disruption and entrainment of jets. I also review several non-linear processes that can have an 
effect on the collimation of extragalactic and microquasar jets. Within those, I remark possible causes for the 
decollimation and deceleration of FRI jets, as opposed to the collimated FRII's. Finally, I give a summary of the 
main reasons why jets can propagate through such long distances.
     
\end{abstract}

\ccode{PACS numbers: 11.25.Hf, 123.1K}

\section{Introduction}	
 Relativistic jets in AGN and microquasars carry energy from very small to very large scales. These jets form in 
the surroundings of compact objects, such as neutron stars or stellar black-holes in the case of microquasar 
jets, and supermassive black-holes (SMBH) in the case of AGN jets. The forming scales are of the order of a few 
times to about 30 times the radius of the central object, depending on the forming mechanism.\cite{bz77,bp82} From these regions, jets propagate 
up to nine powers of ten in distance. This implies formidable stability and collimation taking into account that 
different instabilities can affect the evolution of jets independently of their nature. Moreover, their propagation is 
possibly changing the properties and evolution of the host galaxies and their environments through heating by 
shocks and/or mixing, and removing of gas by shocks and transfer of momentum. Thus, jet stability has been a 
matter of study during the last four decades.\cite{sch76,br76,fe78,ha79}
 
 Depending on the nature of jets, i.e., whether they are magnetically or particle dominated, different types 
of instabilities can arise. In the case of magnetically dominated jets, both current-driven (CDI) and 
Kelvin-Helmholtz (KHI) instabilities can grow, whereas in the case of particle dominated jets, only the latter 
may be present.\cite{ha11} 
 
 Different stabilising mechanisms for both kinds of instabilities have been proposed, and some of them could explain the
remarkable collimation and long distances covered by extragalactic jets, as opposed to the view that the reason for the 
stability of jets remains unknown.\cite{mb09}
 
  In this text I review the state-of-the-art of relativistic jet stability.\footnote{This text will
be restricted to relativistic jets, see the recent review by P.~Hardee\cite{ha11} for a broader view including 
studies for classical jets} Section~\ref{cd} is focused on the CDI. Section~\ref{kh} is devoted to the KHI. 
In Section~\ref{nl} I overview several non-linear processes that can also dramatically affect the evolution of jets and their 
collimation. Finally, in Section~\ref{pv} I discuss the possible reasons why relativistic 
jets manage to carry large amounts of energy from the central regions of galaxies or binary stars to 
very large distances. 

\section{The sub-parsec scales: Current-Driven instability}\label{cd}

Relativistic jets are thought to be formed by magnetohydrodynamical processes in the surroundings of compact 
objects.\cite{bz77,bp82} Thus, it is reasonable to expect that jets are magnetically dominated close to these 
regions.\cite{ha11} In this regime, CDI is triggered by differences in the magnetic forces when a toroidal field
is present. In the case that a pinch is produced, the larger force produced by the compressed lines allows the pinch to grow. 
In the case that a kink is produced, the magnetic lines get closer in the inner part of the kink, increasing the 
magnetic force and enlarging the curvature of the jet. Both modes are stabilised by the presence of a poloidal
field component, as the magnetic tension of these lines acts against the compression or distortion of the jet.

CDI in magnetized flows has been studied from an analytical perspective in the non-relativistic regime by a 
number of authors.\cite{ha11} Results show that this instability becomes dominant at large values of helicity (ratio 
between the toroidal and poloidal field components) 
and it is stabilised by a strong poloidal magnetic field. Under the physical conditions typical of jets close to 
the forming regions, CDI is expected to be dominant, though with small growth rates.\cite{ha11}
In the relativistic regime, numerical simulations of static columns have shown that growth rates decrease with
decreasing Alfv\'en speed and, regarding the non-linear regime, an increasing helicity of the magnetic field with 
increasing radius in the jet, stabilises the flow with respect to kink CDI.\cite{mi09} The introduction of a sub-Alfv\'enic
velocity shear generates important differences in the growth of this instability: In the case that the shear is 
inside the characteristic radius of the static column, the plasma flows through a temporally growing kink, whereas
if the shear is outside that radius, the kink is advected with the flow and grows in distance.\cite{mi11}  

It has been suggested 
that the conversion from Poynting to kinetic flux should occur in the first hundreds of gravitational radii, on 
the basis of observations and modelling of the jet in M87.\cite{ha11,si05,he11} Magnetic 
acceleration has been proposed\cite{ko09} as an efficient mechanism. The growth of CDI to non-linear amplitudes 
can result in mass-load and acceleration of 
the entrained particles, so this process could also be related to a change from magnetically to particle 
dominated jets.\cite{ha11} However, there is no direct evidence for any of both. The role of CDI could
be significantly reduced by jet expansion and rotation, which can have a stabilising effect. 
Further simulations with realistic conditions are needed to solve this question. Nevertheless, extragalactic jets embed stars and are 
necessarily entrained by gas clouds and stars rotating around the AGN, 
so the mass-load and conversion of jet main energy channel from Poynting flux to particles is difficult to avoid. 
It has been shown that sub-Alfv\'enic flows are KHI stable,\cite{mi07} but once the jets are super-Alfv\'enic and 
particle dominated, KHI may take over CDI and start to play an important role in the long-term stability of jets.

\section{The parsec scales: Kelvin-Helmholtz instability}\label{kh}
 The KHI develops at the boundary between two fluids with relative velocity. It couples to any 
perturbation with a certain periodicity at which the system presents unstable modes. The sound-wave generated by 
the perturbation propagates through the body of the jet and grows in amplitude due to over-reflection at the 
boundaries.\cite{pc85} In the non-linear regime, the jet can be disrupted and decelerated 
by the entrainment of external matter. 
 
 Assuming cylindrical symmetry in jets, solutions to the linearised equations of relativistic hydrodynamics have the form 
$\exp(\,i({\bf k}\,{\bf r} + n\,\phi-\omega\,t))$,\cite{ha00} ${\mathbf k}$ 
being the wave-number, $n$ the azimuthal wave-number (an integer giving the number of oscillations around the jet's circular 
cross-section), and $\omega$ the frequency. These solutions may have 
complex values of wave-number and/or frequency, the imaginary part giving the exponential growth of the amplitude. Unstable 
modes are separated into surface modes and body modes. Their distinctive property is the number of 
zeros that the the radial component of the wave-number ($k_r$) of the unstable wave (e.g., the pressure wave) 
has between the jet axis and its boundary. 
The surface mode shows no zeros between the axis and the boundary, whereas the body modes show as many zeros as their order 
indicates (e.g., one zero in the case of the first body mode). Figure~1 shows some extracted modes from the solution 
to the linear problem of a sheared jet. The upper lines represent the real part of the solution (frequency) versus the 
wave-number, whereas the lower lines represent the imaginary part (growth rate) versus wave-number. 

\begin{figure}\label{f1}
\centering{\psfig{trim=0cm 0.5cm 0cm 1cm,clip,file=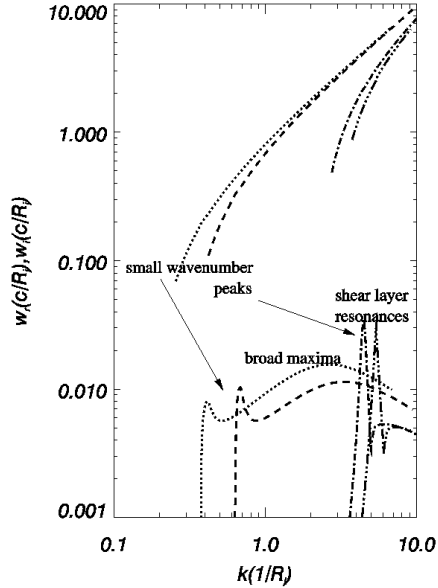,width=0.48\textwidth}} 
 \vspace*{8pt}
 \caption{Selected modes of a solution for a jet with a shear-layer. The plot shows frequency (real part of the solution, upper lines) and 
growth rates (imaginary part of the solution, lower lines) versus
wave-number in units of speed of light and jet radius. Lines with the same format correspond to the real and imaginary part of the selected mode. 
Resonant modes show larger growth rates. From ref.~33.}
 \end{figure}

\paragraph{The linear regime}
 The growth of KHI has been shown to depend on the jet velocity (unstable modes grow faster for slower jets), 
temperature (faster growth for hotter jets) and density (faster growth for more dilute jets).\cite{sch76,br76,fe78,bi91} 
There is a general correlation between this dependence and the long-term stability properties of jets, 
which is more remarkable in colder, faster and denser ones.\cite{pe05}

 The analytical, and numerical in some cases, knowledge acquired about the effect that the growth 
of KHI modes could have on 
jets\cite{ha00,ha87} has been widely used in analysing the structures observed in different extragalactic jets and has allowed to
derive estimates of the physical parameters that govern them (e.g., 3C~345, 3C~120, 3C~273, M87, S5~0836+710).
\cite{he11,ha87}\cdash\cite{pe12} In particular, it was shown that expansion and acceleration
can provide the jet in 3C~120 with long-term stability\cite{ha05} or that the jet in M87 could be decelerating and heating,
leading to destabilization in the kpc scales.\cite{he11} Numerical plus analytical work has also led to the conclusion
that the observed structures in the jet in 3C~273 could be coupling to KHI modes, which may show up at different
observing frequencies, depending on the section of the jet where they develop.\cite{pe06}  This result also has the 
implicit conclusion that different observing frequencies may be showing different regions of a transversally structured jet.
\cite{pl07,pe12} In this respect, several works have also shown that shear-layers are easily generated 
in jets by different mechanisms, including the growth of KHI modes.\cite{pe05,al00}

 In the classical regime, there is no limit to the growth of KHI, other than disruption of the jet or the saturation
of short wavelengths (the latter will be discussed in the section about the non-linear phase), both involving an increase of the width of 
the velocity shear with a decrease of the radial gradient of 
velocity.\cite{bo94} For the case of relativistic jets, there is an analytical prediction of
the saturation of the linear growth when the amplitude of the velocity perturbation reached the speed of light 
in the 
reference frame of the jet.\cite{ha95,ha97} 
Figure~2 shows the saturation of the perturbation in velocity close to the speed of light.
Numerical simulations have confirmed this result and shown that the limiting amplitude is smaller for 
faster jets, thus making them more stable.\cite{pe04a,pe04b} Overall, this shows that relativistic jets are 
intrinsically more stable than classical ones with respect to the growth of any kind of instability.  

 On top of this, semi-analytic plus numerical studies of sheared jets have also shown that the growth of 
small-wavelength, fast-growing modes, as those shown on the right of Fig.~1,\cite{pe05,pe07} can lead to long-term 
collimation of fast jets with moderately thick 
shear-layers ($\simeq 1\,\rm{R_j}$, with $\rm{R_j}$ the jet radius). 

 Different analytical and numerical works have certified that the presence of winds surrounding the 
jet\cite{hh03} or thick shear-layers\cite{ur02,pl07} reduce the growth rates of the KHI modes. It has also been 
shown analytically that in the presence of a poloidal magnetic field, sub-Alfv\'enic and even super-Alfv\'enic jets 
can be stabilised by the presence of a magnetised surrounding sheath.\cite{ha07} 

\begin{figure}\label{f2}
\centering{\psfig{trim=0cm 0.4cm 0cm 0.8cm,clip,file=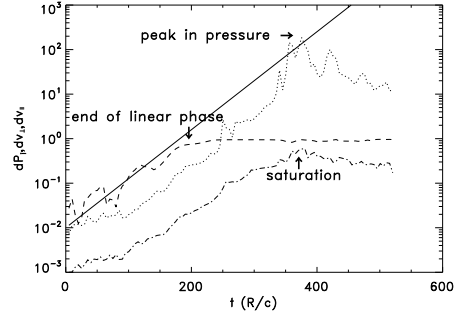,width=0.50\textwidth}}
 \vspace*{8pt}
 \caption{Plot of the linear growth of KHI modes in a relativistic flow versus time. The dotted line represents the amplitude of the 
pressure perturbation and the dashed and dash-dotted lines represent the amplitude of the perturbation of the axial and radial
components of velocity, respectively, in the reference frame of the jet. The units are speed of light, jet radius and ambient medium density. 
From ref.~31.}
 \end{figure}

\paragraph{The non-linear regime}
 starts typically after saturation.\cite{pe04a,pe04b} If the dominating mode is a 
low-order one with a relatively long wavelength, the jet shows strong deformations and can be disrupted by the 
generation of shocks at the boundary with the ambient medium.\cite{pe05,pe04b} In this context, the jet develops a wide 
mixing layer and undergoes significant deceleration to mildly relativistic speeds. Otherwise, if small-wavelength modes like 
resonant modes (see Fig.~1) dominate the growth of the instability, the jet develops a hot shear-layer and keeps collimation.\cite{pe05,pe07} 
Figure~3 shows the Schlieren plots for the last snapshot of two simulations of the temporal growth of KHI in jets with different 
Lorentz factors ($\gamma=5$ for the left image and $\gamma=20$ for the right one). Otherwise, both simulations are equal. 
The left panel, corresponding to a slower jet developing long-wavelength modes, shows turbulent mixing in a wide region, whereas 
the right panel, corresponding to the faster jet, which develops resonant modes, shows a significant degree of collimation.
This result has been recently confirmed by 3D simulations and its implications extensively discussed\cite{pe10} (see Figure~4).
A similar, although not so efficient, mechanism for long-term jet stabilisation was also found for classical jets when the shortest modes 
dominate the growth of the instability up to the non-linear regime,\cite{xu00} which is also true for relativistic jets.\cite{pe04b} 

 The distribution of a number of simulated jets in a 
relativistic Mach-number versus Lorentz factor plane\cite{pe05,pe04a,pe04b} shows that there is a trend of larger stability in 
the non-linear regime in the case of faster and colder jets, which fall in the upper-right corner of the plane. 

 \begin{figure}\label{f3}
\psfig{file=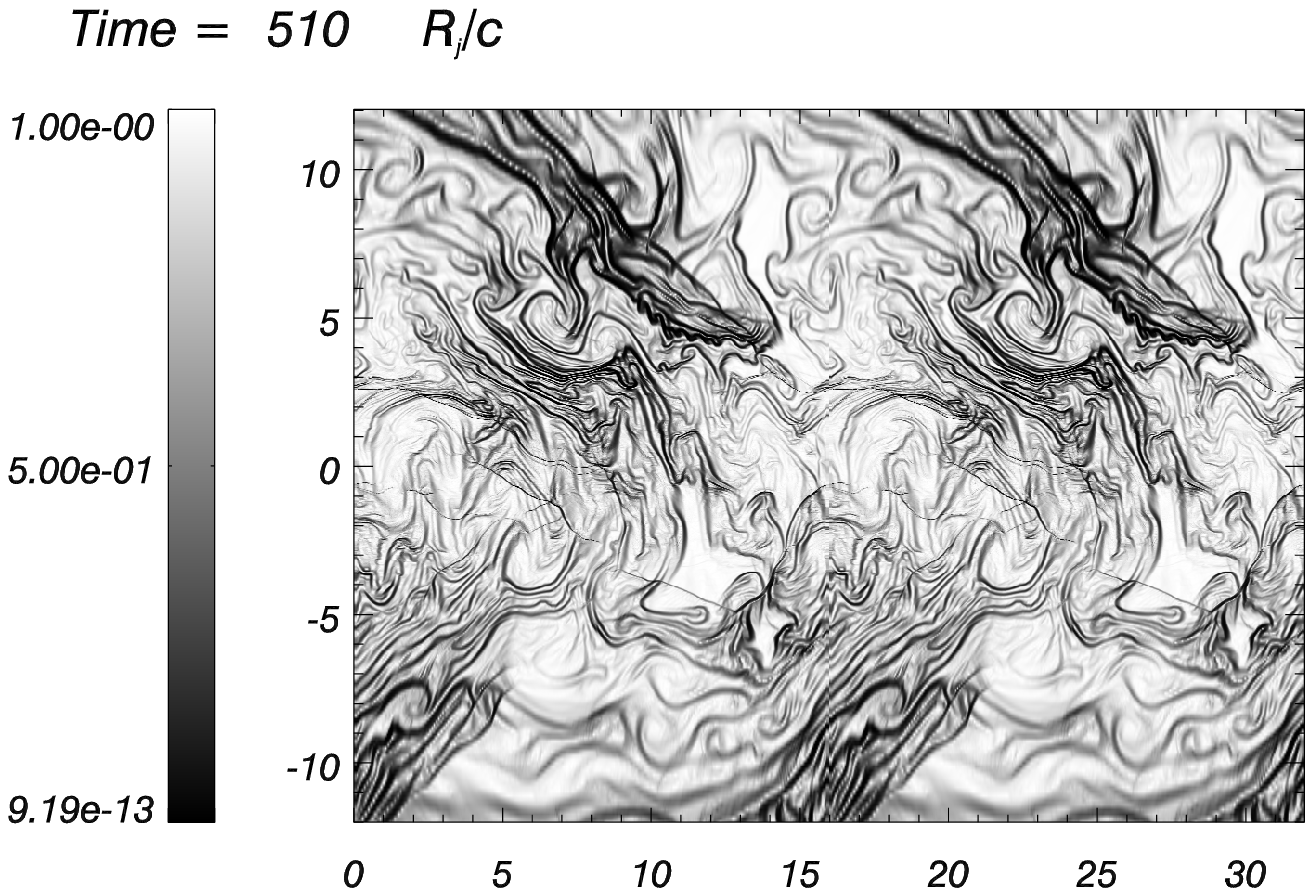,width=0.48\textwidth}
\psfig{file=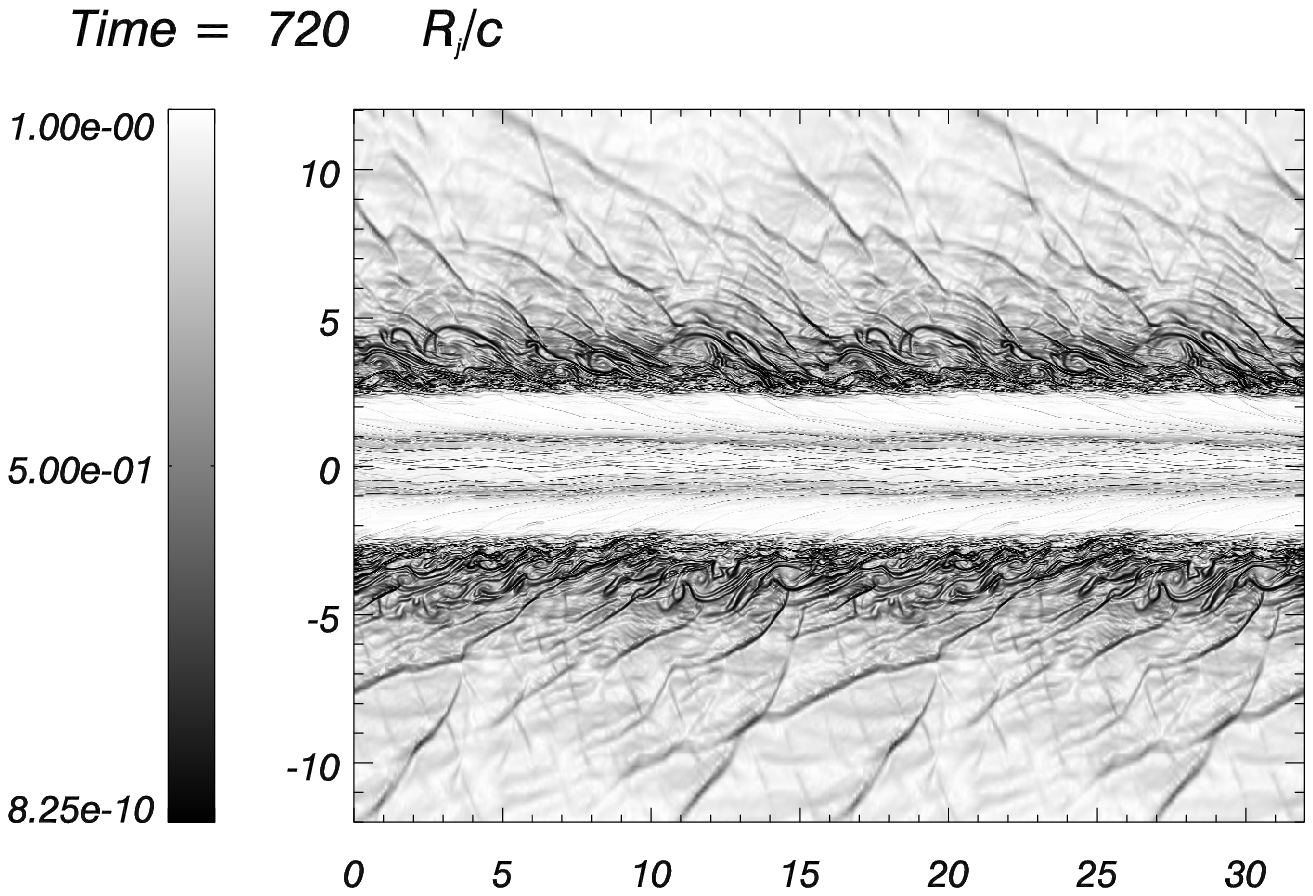,width=0.48\textwidth}
 \vspace*{8pt}
 \caption{Schlieren (enhanced density gradients) plots of two jets, one developing long wavelength modes (left) and another developing resonant
(short-wavelength) modes. Axis are jet radii. From ref.~33.}
 \end{figure}

\begin{figure}\label{f4}
\centering{\psfig{file=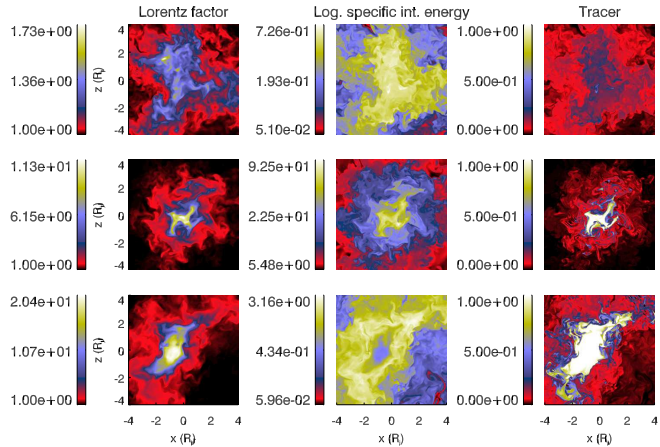,width=0.78\textwidth}}
 \vspace*{8pt}
 \caption{Transversal cuts from 3D simulations of a cold jet with Lorentz factor 5 (top row) and 20 (bottom row) and an hot jet with 
Lorentz factor 10 (middle row). The cuts show Lorentz factor, logarithm of the specific internal energy (units of $c^2$) and jet mass 
fraction wrt original, from left to right. The faster jet shows ($\gamma=20$, bottom row) less mixing, preserves a larger Lorentz factor and develops a 
hot shear-layer, as predicted 
by the 2D simulations. The slow jet ($\gamma=5$, top row) undergoes strong mixing and deceleration, whereas the hot jet ($\gamma=10$, top row)
shares properties with both previous cases. From ref.~37.}
 \end{figure}

\section{Non-linear effects}\label{nl}

 In the previous section, the growth of instabilities has been discussed as starting in the linear regime, i.e., 
from small perturbations. However, jets are destabilised by perturbations with large (non-linear) amplitudes 
and may undergo reconfinement shocks or meet irregularities 
in the ambient media. These processes are non-linear and cannot be accounted for by linear theory of jet 
stability. They are of interest because generally imply strong shocks and thus represent candidate 
locations for very-high energy emission. 

\paragraph{Extragalactic jets} Most probably jets are not in pressure equilibrium with the ambient medium
because the pressure in the surrounding cocoon changes with time, as it expands. Over-pressured jets 
expand and recollimate after becoming under-pressured with respect to the environment,\cite{dm88,fa91} or meet high density
regions in an inhomogeneous environment (e.g., supernova remnants (SNR) or massive stars with powerful winds). Depending 
on the initial pressure difference, these shocks can trigger small amplitude pinching or helical motion that may couple to KHI,\cite{ag01} 
or they can generate a large amplitude pinch and destroy the jet\cite{pm07} (see Figure~5). Only in the case that jets are close to pressure 
equilibrium with respect to the ambient medium, the generation and influence of reconfinement shocks can be neglected. 

 When crossing the galaxy, the jet flow may encounter stars and clouds of gas that can be embedded in the jet or 
entrain it due to their proper motions.\cite{ar10,ba10} In the case of stars, the stellar wind has been claimed to efficiently 
mix with the jet flow and decrease the mean jet velocity,\cite{ko94,lb02,hb06} mainly in the case of massive stellar 
winds and very light jets. Also, irregularities in the intergalactic medium have been claimed to play a role in the 
hybrid morphology of a number of sources that show both FRI and FRII structure in their two jets.\cite{gw00,mk08} 

 Large amplitude initial perturbations, such as those triggered by changes in the injection angle of the flow could generate 
non-linear distortions of jets and lead to their disruption.\cite{ro08}
The injection of overdense plasma in jets has been invoked to explain the observation of components travelling 
through parsec-scale jets and the generation of trailing components.\cite{ag01,ka08,pe08,pm09} These have been shown to 
generate significant pinching that could couple to KHI modes.\cite{ag01} In the kiloparsec scales, the arcs observed 
in some FRI jets\cite{} have also been suggested to be due to changes in the injection rates from the source.\cite{la08}
 
 \begin{figure}\label{f5}
\centering{\psfig{file=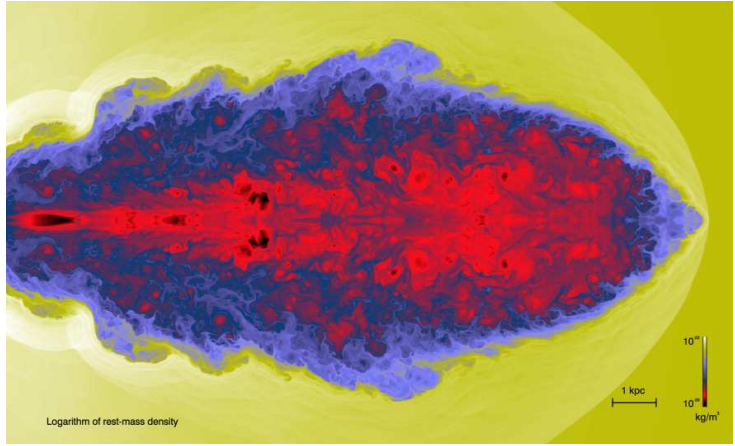,width=0.76\textwidth}}
 \vspace*{8pt}
 \caption{Snapshot of rest-mass density from a long-term 2D simulation of a FRI-like jet evolving in a King-density profile. The jet 
is disrupted after non-linear pinching triggered by a strong reconfinement shock at $\simeq 1.5$~kpc from the injection in the numerical 
grid. The colour bar ranges from $10^{-32}$ to $10^{-25}\,\mathrm{g/cm^3}$. From ref.~42.}
 \end{figure}

\paragraph{Microquasar jets} Inhomogeneities in the ambient medium can also have a strong effect on microquasar 
jets in high-mass X-ray binaries (HMXB) with typical powers ($10^{36-37}\,\rm{erg/s}$).\cite{br11} 
Such inhomogeneities are encountered by the jet in the transition between the unshocked and shocked stellar wind 
from the massive companion. Another change of ambient medium occurs for a young HMXB still embedded 
in the SNR, at the transition between the shocked wind and the shocked SNR, and finally from the latter 
to the interstellar medium (ISM) through the shocked ISM. For the case of older HMXB, the system has moved away from the SNR or this 
has become very dilute and the region of interaction between the shocked stellar wind and the shocked ISM also implies a 
change in the conditions encountered by jets. In these density jumps jets are decelerated 
and need several thousand years to drill the shocked ISM before being able to propagate into the ISM. 
Proper motion of the HMXB 
provides a further destabilising mechanism, as the jet is impacted by the otherwise quasi-steady shocked stellar 
wind, propagating in the direction opposite to that of the proper motion. 

Within the binary system, the stellar wind impacting the jet can also generate jet entrainment, deceleration and loss 
of collimation for jets with powers $\leq 10^{37}\,\rm{erg/s}$, mainly if a strong reconfinement shocks that decelerate 
the flow occur in this region, as has been shown to be the possible.\cite{pb08,pbk10} See also Figure~6.  

 \begin{figure}\label{f6}
\centering{\psfig{trim=0cm 3.5cm 0cm 2cm,clip,file=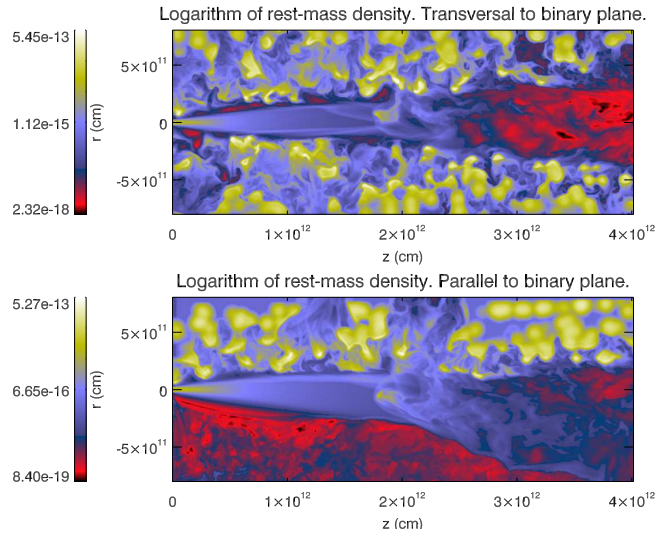,width=0.8\textwidth}}
 \vspace*{8pt}
 \caption{Two perpendicular cuts along the axial direction of rest-mass density from a 3D simulation of a jet interacting with an 
inhomogeneous stellar wind in a HMXB. Units are cgs. From ref.~57.}
 \end{figure}

  

\section{A global (and personal) view on the stability of relativistic jets}\label{pv}

 In the previous sections, the main properties of instabilities that can affect the large-scale morphology of relativistic 
jets, namely CDI and KHI, have been summarised. 
These have been proven to be disruptive enough to trigger efficient entrainment and decelerate jets. However, 
several mechanisms that may damp their growth and thus make relativistic jets stable have been given:
\begin{itemize}
 \item CDI: Poloidal magnetic field, mildly relativistic sheaths and jet expansion.
 \item KHI: Thick shear-layers or surrounding winds, the decrease of the cocoon density with time and jet expansion.
\end{itemize}

  Most importantly, it has been shown that relativistic jets are less sensitive to the growth of instabilities than classical jets, due to 
the saturation of their growth at the speed of light. In addition, resonant modes that can show up in fast jets surrounded by thin shear-layers, 
or simply the development of short wavelength modes as compared to the jet radius can result in little mixing, restricted to the boundaries of
jets, and thus little loss of jet thrust and collimation. If one or several of these mechanisms play a role, jets can remain collimated along 
large distances, as it has been shown by simulations. Other mechanisms such as jet rotation or different configurations of the magnetic field 
are still to be explored in the relativistic regime. 

  The properties of jets and the media outside the leading bow-shock change with distance to the source. The cocoons 
surrounding jets show a homogeneous pressure in long-term simulations of jets due to their sound-speeds being larger than the 
expansion velocities. 
As the cocoon expands, its density and pressure drop and, as long as the properties of the jet do not change, it will become 
denser with respect to its environment, gaining extra stability. The only caveat is that when the cocoon is very 
under-pressured, strong reconfinement shocks may form. Nevertheless, even when jets are entrained and decollimated, they seem to preserve 
a large velocity that can be sub-relativistic, but could still be enough to keep the propagation of the jet to large scales.

 In FRI jets, as compared to FRII's, it is more likely that non-linear processes like mass-loading from stars\cite{lb02} or 
strong reconfinement shocks\cite{pm07} are the cause of their disrupted morphologies. These jets have apparently similar 
velocities in the parsec-scales as the FRII's,\cite{gio01} so other intrinsic properties such as density, temperature or 
composition should 
play a role to explain their lower powers. They are observed to have an opening angle in the inner regions, but there is not any hint of 
growth of instabilities before the jet deceleration and loss of collimation. In hybrid sources in which the jet and counterjet
appear to have different morphologies,\cite{gw00} it is possible that irregularities in the ambient medium are 
the cause of the difference, for jets with powers between those typical of FRIs and FRIIs.
Only in one case an FRII classified jet with an 
apparently irregular structure in the largest scales,\cite{hu92} has been reported to develop instabilities that could cause 
the jet disruption.\cite{pe12} This implies that, under regular conditions, i.e., without strong non-linear perturbations,  
the stabilising mechanisms proposed by different authors in the last years and summarised here play a significant role.
It has also been suggested that FRII jets can turn into FRI jets in the 
long term\cite{wa11} or after the end of an active phase, as numerical simulations show.\cite{pqm12} However, this possibility should be 
matched to the relativistic velocities measured in FRI jets at the parsec scales.       

The long-term stability of relativistic jets allows them to carry energy from very small to very large scales. 
The bow-shocks generated by extragalactic jets need $10^7 - 10^8$~years to reach distances of the order of 
hundreds of kpc. In the most powerful radio-sources, strong shocks are triggered in the ISM and ICM, which can change the 
evolution of the galaxies themselves by removal of ISM. This process can suppress star formation significantly and heating the surrounding
gas, as shown by recent long-term simulations of relativistic jets.\cite{pqm12} Thus, the importance of the study of this 
interaction and the evolution of the jet itself.

 The research on jet stability in the next years should be addressed to investigate the effect of magnetic fields on the 
long-term evolution of jets. This should help to study, with input from observations and modelling, 
to which distance and extent the magnetic field is dynamically important and the role of instabilities in the eventual 
transfer of Poynting flux into kinetic energy of particles. In addition, more realistic situations should be explored, such 
as jet rotation and expansion. These works will certify or falsify the role of the stabilising mechanisms summarised in this
contribution and could even add further ones, suggesting that there may be more than one answer to the problem of jet stability. 
Finally, it would be important to study, via detailed numerical simulations, 
the process of mixing and entrainment that jets undergo as a consequence not only of the growth of instabilities, but also 
of the interaction with stellar winds and clouds of cold gas. 

\section*{Acknowledgements}
I acknowledge J.M.~Mart\'{\i}, M.~Hanasz and P.E.~Hardee for sharing their understanding of jet physics with me
during the last years.   
Financial support by the Spanish ``Ministerio de Ciencia e Innovaci\'on''
(MICINN) grants AYA2010-21322-C03-01, AYA2010-21097-C03-01 and
CONSOLIDER2007-00050 is also acknowledged.

\end{document}